# Interference Induced Asymmetric Transmission Through A Monolayer of Anisotropic Chiral Metamolecules

Shuang Zhang[1*], Fu Liu[2], Thomas Zentgraf[3], Jensen Li[1, 2*]

1. School of Physics & Astronomy, University of Birmingham, Birmingham, B15 2TT, UK

2. Department of Physics and Materials Science, City University of Hong Kong, Tat Chee Avenue, Kowloon Tong, Hongkong, China

3. Department of Physics, University of Paderborn, Warburger Straße 100, D-33098 Paderborn, Germany

We show that asymmetric transmission for linear polarizations can be easily achieved by a monolayer of anisotropic chiral metamolecules through the constructive and destructive interferences between the contributions from anisotropy and chirality. Our analysis is based on the interaction of electromagnetic waves with the constituent electric and magnetic dipoles of the metamaterials, and an effective medium formulation. In addition, asymmetric transmission in amplitude can be effectively controlled by the interference between spectrally detuned resonances. Our findings shed light on the design of metamaterials for achieving strong asymmetric transmission.

Metamaterials have gained growing interest over the past decade because of many potential applications, ranging from sub-diffraction imaging to invisibility cloaks [1-4]. Negative refractive index and gigantic optical activities are fascinating examples of the exotic electromagnetic properties that arise from the subwavelength structural effect, rather than the constituent materials of the metamaterials [5-9]. Metamaterials can be designed to exhibit symmetries and topologies [10-13] that go beyond natural materials, and hence may introduce entirely new physical effects and electromagnetic phenomena.

Moreover, metamaterials are capable of enhancing the electromagnetic effects arising from strong symmetry breaking, examples include the metamaterial analog of electromagnetically induced transparency [14-16].

It was recently discovered that metamaterials with strong symmetry breaking could exhibit the intriguing phenomenon of asymmetric transmission of light for circular [17-21] and linear polarizations [22-24]. While the asymmetric transmission for circularly polarized waves occurs at planar metamaterials that lacks mirror and rotational symmetry (three fold or above) in the in-plane directions, the asymmetric transmission for linear polarizations requires a more complex 3D metamaterial design that entails mirror symmetry breaking in all three dimensions, indicating that the asymmetric transmission for linear polarizations involves presence of chirality. From the argument based on symmetry [22], one can infer that chirality and anisotropy are the necessary conditions for achieving asymmetric transmission. However, there has been no discussion in a quantitative manner how the anisotropy and chirality contribute to asymmetric transmission for linearly polarized waves. In addition, it is not clear whether additional requirements besides chirality and anisotropy are needed for inducing asymmetric transmission. To this end, a formal analysis based on the interaction of electromagnetic waves with the constituent electric and magnetic dipoles may present more complete information of the origin of asymmetric transmission, and provide a quantitative guidance on the design of metamaterials for maximizing the asymmetry transmissions.

The phenomenon of asymmetric transmission can be described by the Jones matrix, which relates the incident and transmitted wave as,

$$\begin{pmatrix} T_x \\ T_y \end{pmatrix} = \begin{pmatrix} A & B \\ C & D \end{pmatrix} \begin{pmatrix} I_x \\ I_y \end{pmatrix} \qquad (1)$$

where $T_x$, $T_y$ are the transmitted fields in $x$ and $y$ polarizations, $I_x$ and $I_y$ are the fields of the incident wave. Based on reciprocity principle, the Jones matrix is simply transposed for a wave incident from the opposite direction. Thus, the difference in transmission in the opposite directions for linear polarizations can be characterized by the difference between the amplitudes of the two off-diagonal elements, $B$ and $C$, which can be determined from the effective electromagnetic parameters of the metamaterials, including the permittivity and permeability tensors, and the coupling coefficients between the electric and magnetic responses. These effective electromagnetic parameters are fundamentally associated with how light interacts with the constituent electric and magnetic dipoles of the building block (unit cell) of the metamaterials.

In general, a metamaterial unit cell with low symmetry can be modeled as coupled electric and magnetic dipoles. The electric and magnetic dipole moments of the coupled dipoles are related to the electric and magnetic fields of incident light by,

$$\begin{pmatrix} p_x \\ p_y \\ m_x \\ m_y \end{pmatrix} = \begin{pmatrix} \alpha^e_{xx} & \alpha^e_{xy} & \alpha^{EH}_{xx} & \alpha^{EH}_{xy} \\ \alpha^e_{yx} & \alpha^e_{yy} & \alpha^{EH}_{yx} & \alpha^{EH}_{yy} \\ \alpha^{HE}_{xx} & \alpha^{HE}_{xy} & \alpha^m_{xx} & \alpha^m_{xy} \\ \alpha^{HE}_{yx} & \alpha^{HE}_{yy} & \alpha^m_{yx} & \alpha^{HE}_{yy} \end{pmatrix} \begin{pmatrix} E_x \\ E_y \\ B_x \\ B_y \end{pmatrix} \quad (2)$$

where $p_x, p_y, m_x, m_y$ are the x- and y- components of the electric and magnetic dipoles, $\alpha^e$, $\alpha^m$ are the electric and magnetic dipole moments, $\alpha^{EH}$ and $\alpha^{HE}$ are the coupling strength between the dipoles, with $\alpha^{HE}_{ij} = -\alpha^{HE}_{ji} = \alpha^{EH}_{ij} = -\alpha^{EH}_{ji}$ and $\alpha^{e(m)}_{ij} = \alpha^{e(m)}_{ji}$.

We consider an x-polarized electromagnetic wave propagating along +z (-z) direction. The corresponding electric and magnetic fields are given as ($E_0 \hat{e}_x, \pm B_0 \hat{e}_y$), where $B_0 = E_0/c$. The y polarized forward radiation is generated from $m_x$ and $p_y$, which are given as,

$$m_x = \pm \alpha^m_{xy} B_0 - \alpha^{EH}_{xx} E_0$$

$$p_y = \alpha^e_{xy} E_0 \pm \alpha^{EH}_{yy} B_0$$

The corresponding radiation field in the forward direction for an incident beam in the ±z directions is given as,

$$E^\pm_y = \{-\frac{1}{4\pi}\sqrt{\frac{\mu_0}{\varepsilon_0}} k^2 (\pm \hat{z} \times m_x \hat{x}) + \frac{1}{4\pi\varepsilon_0} k^2 (\pm \hat{z} \times p_y \hat{y}) \times (\pm \hat{z})\} \frac{e^{ikz}}{z}$$
$$= \frac{1}{4\pi}\sqrt{\frac{\mu_0}{\varepsilon_0}} k^2 \frac{e^{ikz}}{z} [\alpha^e_{xy} c - \alpha^m_{xy}/c \pm (\alpha^{EH}_{xx} + \alpha^{EH}_{yy})] E_0 \quad (3)$$

Eq. (3) shows that the contribution to cross polarized radiation consists of two symmetry related terms, anisotropy ($\alpha_{xy}^e c - \alpha_{xy}^m/c$) and chirality ($\alpha_{xx}^{EH} + \alpha_{yy}^{EH}$), and the difference between $E^+$ and $E^-$ arises from the constructive (+) and destructive (-) interferences between them. We wrote Eq. (3) in terms of dipolar response for clarity of explanation. For an array of metamolecules, the radiated plane waves are simply the summation of the radiation from each individual dipole. Eq. (3) provides a mathematical proof that both anisotropy and chirality are required for inducing asymmetric transmission. As the chirality term is the trace of a 2×2 tensor in the x-y dimension, it is invariant under the rotation of the coordinate in the x-y plane. In other words, the contribution to the cross polarization radiation from the chiral term is independent of the polarization of the incident wave.

Now, let us consider a metamaterial unit cell consisting of an electric dipole and a magnetic dipole coupled to each other and forming an angle $\psi$, as shown in Fig. 1(a). For this configuration, the forward scattering in Eq.. (3) can be rewritten as,

$$E_y^\pm \propto \frac{\alpha^e}{2} c \sin 2\theta - \frac{\alpha^m}{2c} \sin 2(\theta+\psi) \pm \alpha^{EH} \cos\psi \quad (4)$$

where $\theta$ and $\phi = \theta + \psi$ are the angles formed by the electric and magnet dipole with the x-axis, respectively. Eq. (4) shows that the difference between $E_y^+$ and $E_y^-$ vanishes only for $\psi = \pi/2$, i.e. the electric and magnetic dipoles are orthogonal to each other, and in this particular case the resonator becomes purely bianisotropic without chirality.

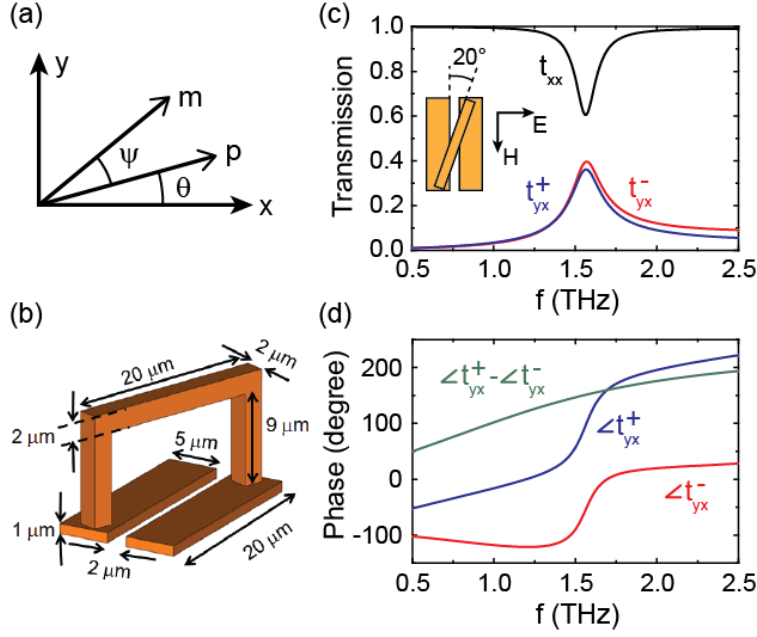

**Fig. 1**. **(a)** A chiral structure can be modeled as coupled electric and magnetic dipoles, forming an angle of θ and ϕ, respectively, with the x-axis. **(b)** Schematic of a chiral split ring resonator (SRR) made of gold, with the geometric parameters indicated in the figure. **(c)** The transmission spectra, $t_{xx}$ (black) and $t_{yx}^{+}$ (blue), and $t_{yx}^{-}$ (red), for a metamaterial consisting of an array of chiral SRRs. **(d)** the phase of transmission for $t_{yx}^{+}$ (blue) and $t_{yx}^{-}$ (red), and their difference (green). The inset shows a top view of chiral SRR. The structure is arranged periodically with equal period of 30 μm in both directions.

For a meta-atom exhibiting a single resonance with coupled electric and magnetic responses, the electric and magnetic polarizabilities and the coupling coefficients take the forms of [25],

$$\alpha^e = \frac{A_e \omega_0^2}{\omega_0^2 - \omega^2 - i\gamma\omega}, \quad \alpha^m = \frac{A_m \omega^2}{\omega_0^2 - \omega^2 - i\gamma\omega}$$

$$\alpha^{EH} = \pm j \frac{\sqrt{A_e A_m}\, \omega_0 \omega}{\omega_0^2 - \omega^2 - i\gamma\omega} \tag{5}$$

It follows from Eq. (5) that there exists a π/2 phase difference between $\alpha^{m(e)}$ and the coupling term $\alpha^{EH}$. Therefore, according to Eq. (4), the difference between $E_x^{+}$ and $E_x^{-}$ is manifested in phase rather than in amplitude. This is confirmed by numerical simulation on a metamaterial whose unit cell consists of a twisted chiral SRR, as shown in Fig. 1 (b).

This chiral metamaterial design has recently led to the demonstration of chirality induced negative refractive index [8]. The structure can be considered as an LC resonator, wherein the metal loop functions as an inductor and the gap between the two base metal strips function as a capacitor. The meta-atom exhibits a strong chirality at the LC resonance frequency at 1.57 THz, where the transmission spectra exhibit pronounced features: a resonance dip in $t_{xx}$ and a resonance peak in $t_{yx}^{+}$ and $t_{yx}^{-}$. Interestingly, although the amplitudes of $t_{yx}$ in the +z and -z directions are almost identical [Fig. 1(c)], their phases are dramatically different, as shown in Fig. 1(d).

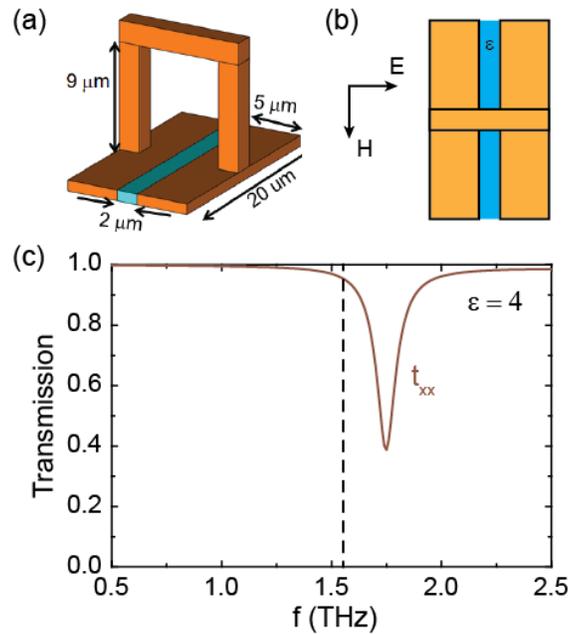

**Fig. 2**. **(a, b)** Schematics of the metamaterial unit cell consisting of an achiral SRR (R2). The gap of the achiral SRR is filled with a material with dielectric constant $\varepsilon = 4$. **(c)** The transmission spectrum $t_{xx}$ of the achiral metamaterial. The dashed line indicates the resonance frequency of the chiral resonator as in Fig. 1(c). In the simulation, the unit cell is the same as that shown in Fig. 1(b).

To induce asymmetric transmission in amplitude, contribution from additional electric or/and magnetic dipole moments of different phases from that of the chiral resonance is required. This can be realized by adding another resonant structure with resonance frequency slightly detuned from that of the chiral resonator to the metamaterial unit cell. A specific design of the additional resonator (R2) is shown in the inset of Fig.

2(a, b): a SRR with mirror symmetry, where the electric dipole (across the gap) and the magnetic dipole (perpendicular to the loop) are forming an angle $\psi_2 = \pi/2$. The geometry of R2 and the dielectric constant of the material filling the gap are designed such that the resonance frequency of R2 is slightly blue-shifted from that of the chiral resonator, as shown in Fig. 2(c).

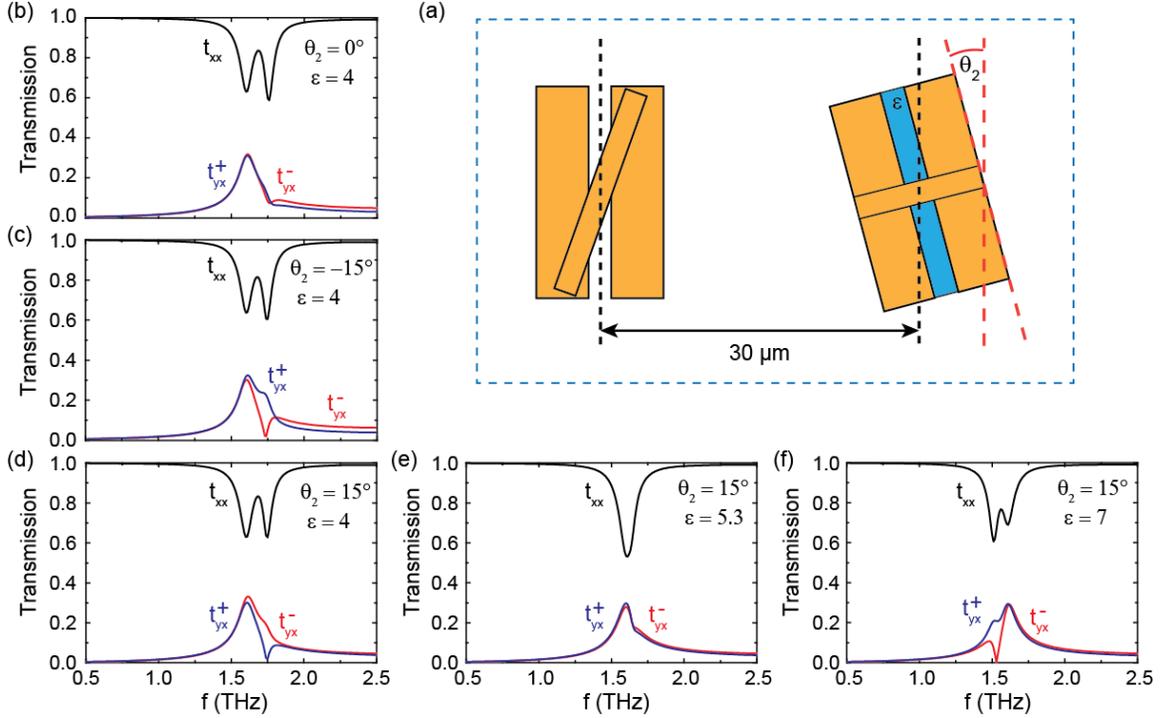

**Fig. 3**. **(a)** Schematic of a metamaterial unit cell consisting of both the chiral and achiral SRRs. The size of unit cell is 60 μm × 30 μm. The orientation of the achiral SRR is indicated by the angle $\theta_2$ formed between the base metal strips and the y-axis. **(b, c, d)** Transmission spectra for $t_{xx}$ (black), $t_{yx}^{+}$ (blue) and $t_{yx}^{-}$ (red) at different orientations of the achiral element $\theta_2 = 0°$, -15°, +15°, while the permittivity of the gap material of R2 [blue area in panel (a)] is fixed at $\varepsilon = 4$. **(e, f)** Transmission spectra for different gap material permittivities $\varepsilon = 5.3$ and 7, respectively, while the orientation of the achiral element is fixed at $\theta_2 = 15°$.

The schematic of the composite metamolecule consisting of the chiral resonator and the achiral one (R2) is shown in Fig. 3(a). For an x-polarized incident plane wave, the forward scattering into orthogonal (y) polarization by the metamolecule is expressed as,

$$E_y^\pm \propto \frac{\alpha_1^e c}{2}\sin 2\theta_1 - \frac{\alpha_1^m}{2c}\sin 2(\theta_1 + \psi_1) + \left(\frac{\alpha_2^e c}{2} + \frac{\alpha_2^m}{2c}\right)\sin 2\theta_2 \pm \alpha_1^{EH}\cos\psi_1 \qquad (6)$$

where indices '1' and '2' refer to the chiral and achiral resonators (R2), respectively. Note that the contribution from R2 to the forward scattering can be tuned by varying its orientation angle $\theta_2$ and its resonance frequency.

The full wave simulation results on the transmission of electromagnetic waves through a metamaterial made of the composite metamolecules are shown in Fig. 3(b-d) at various orientation angles of R2. For all the orientations, $t_{xx}$ spectrum (black) shows two resonance dips at 1.61 THz and 1.75 THz, corresponding to the resonance frequencies of the chiral and achiral element, respectively. Note that the presence of coupling between the chiral and achiral elements is indicated by a slight resonance shift in $t_{xx}$ relative to that of the uncoupled case. At $\theta_2=0$, the transmission spectra in the forward and backward direction are nearly identical (Fig. 3(b)) because there is no contribution from R2 to the y-polarized radiation. A slight rotation in either counter clockwise ($\theta_2=15°$) or clockwise direction ($\theta_2=-15°$) introduces strong asymmetric transmission between propagation along +z and −z directions, as shown in Fig. 3(c) and (d). Despite the almost identical spectral responses in $t_{xx}$ for $\theta_2=\pm15°$, their cross polarized transmissions ($t_{yx}$) show dramatic difference. For $\theta_2=-15°$ (Fig. 3c), there appears a destructive interference in $t_{yx}^-$ between the contributions from the chiral and achiral elements, leading to a sharp dip approaching zero at the resonance frequency of R2 (red). On the other hand, for +z incidence, the interference is constructive, as indicated by a hump in $t_{yx}^+$ at the resonance frequency of R2 (blue). Interestingly, for $\theta_2=+15°$, the asymmetry is reversed: the wave incident along -z exhibits much larger transmission of cross polarization than that of +z incidence at the resonance frequency of R2. The above observations are in good agreement with Eq. (6), which shows that rotation of R2 along opposite directions (flipping the sign of $\theta_2$) gives rise to opposite asymmetric transmissions.

We next study the effect of the detuning of the resonance frequencies between the chiral and achiral elements on asymmetric transmission. In the simulation, the resonance frequency of the achiral element is tuned by varying the dielectric constant $\varepsilon$ of the dielectric material in the gap region. The orientation of the achiral element is fixed at $\theta_2=15°$, where a large asymmetric transmission with $t_{yx}^+ < t_{yx}^-$ has been shown for $\varepsilon=4$ as in Fig. 3(d). With the increase of the permittivity of the gap material, the resonance frequency of the achiral element decreases and crosses that of the chiral element at a permittivity around $\varepsilon$ =5.3, where the two resonance dips in $t_{xx}$ merge into one [Fig. 3(e)]. Interestingly, despite the low symmetry in the structure, there is almost no asymmetric transmission as the two cross polarization transmission spectra along opposite directions are almost identical. This can be attributed to the fact that the electric and magnetic dipoles of the achiral elements are in phase with that of chiral elements, and according to Eq. (6), there should be no difference in the transmission amplitudes in the two opposite incident directions.

With further increase of the dielectric constant to 7, the resonance of the achiral element R2 is shifted to a lower frequency of 1.52 THz, which falls below that of the chiral element (~1.6 THz). Fig. 3(f) shows a strong asymmetry in transmissions around the resonance frequency of R2. In contrast to the results of low dielectric gap material in Fig. 3(d), the cross polarization transmission spectra in Fig. 3(f) shows the opposite asymmetry, namely, light propagating along -z direction exhibits a pronounced dip in the vicinity of the resonance frequency of R2, resulting in $t_{yx}^- < t_{yx}^+$. This is of no surprise, as sweeping the resonance frequency of the achiral resonator across that of the chiral one changes the sign of the relative phase between them, which, according to Eq. (6), leads to reversed asymmetry in transmission for incident wave propagating along opposite directions.

For a metamaterial consisting of an array of chiral atoms, its effective parameters are the macroscopic representation of the polarizabilities of individual chiral atom, and therefore the relationship between the cross polarized transmission for a linearly polarized wave and its effective parameters can be formulated in a similar manner as Eq. (3). Through a rigorous derivation for a thin slab composed of a plane of dipoles (See Appendix for further details.), it is shown that the ratio between the cross polarized transmission along opposite directions, $t_{yx}^+$, $t_{yx}^-$, is related to the effective parameters in the following approximation:

$$\frac{t_{yx}^+}{t_{yx}^-} \approx \frac{(\varepsilon_{xy} - \mu_{xy}) + (\xi_{xx} + \xi_{yy})}{(\varepsilon_{xy} - \mu_{xy}) - (\xi_{xx} + \xi_{yy})} + O(d) \qquad (7)$$

where $\varepsilon_{ij}$, $\mu_{ij}$ and $\xi_{ij}$ denote the effective permittivity, permeability and chiral tensor elements, respectively. Note that there is a correspondence between the effective parameters in Eq. (7) and the polarizability terms in Eq. (3), with $\varepsilon_{xy} \propto \alpha^{e}_{xy} c$, $\mu_{xy} \propto \alpha^{m}_{xy}/c$, and $\xi_{ii} \propto \alpha^{EH}_{ii}$. From Eq. (7) follows that asymmetric transmission for linear polarization from a planar array of dipoles can only be obtained with anisotropy ($\varepsilon_{xy}$ or $\mu_{xy}$) and chirality ($\xi_{xx}$ or $\xi_{yy}$) in order to make a large difference between the numerator and denominator. Moreover, loss will be needed otherwise $\varepsilon_{xy} - \mu_{xy}$ becomes purely real and $\xi_{xx} + \xi_{yy}$ becomes purely imaginary to give asymmetric transmission in phase but not in amplitude.

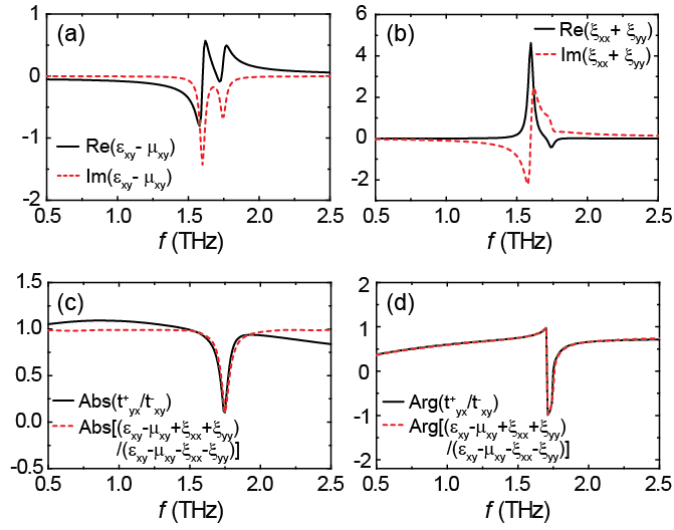

**Fig. 4.** **(a)** The real and imaginary part of the anisotropy term ($\varepsilon_{xy} - \mu_{xy}$). **(b)** The real and imaginary part of the chiral term ($\xi_{xx} + \xi_{yy}$) **(c, d)** The amplitude and phase of the ratio between $t_{yx}^{+}$ and $t_{yx}^{-}$ calculated by simulation (black) and by Eq. (7) (red).

Equation (7) is numerically validated on a metamaterial with the same unit cell shown in Fig. 3(a) with $\theta_2=15°$ and $\varepsilon=4$. We have retrieved the polarizabilities and the effective parameters of the asymmetric metamaterials numerically from the complex coefficients of the transmission and reflectance (see appendix for details). As shown in Fig. 4(a), the retrieved anisotropy term exhibits two dips at resonance frequencies of the chiral and achiral elements, as both resonators are anisotropic structures. On the other hand, the chirality term, as plotted in Fig. 4(b), shows a pronounced resonance peak only at the resonance frequency of the chiral resonator. As shown by Fig. 4(c) and 4(d), the ratio between transmissions along opposite directions, shows almost perfect match to that given by Eq. (7), in both amplitude and phase. At the resonance frequency (1.61 THz) of the chiral element, despite the strong anisotropy and chirality, there is no noticeable asymmetric transmission at this frequency. Instead, the ratio of the amplitudes exhibits a sharp dip approaching zero at the resonance of the achiral element around 1.75 THz.

In conclusion, by using metamaterial as a platform, we have investigated the influence of multiple types of broken symmetries over optical functionality. In particular, we showed how the constructive and destructive interferences between two types of broken symmetries: anisotropy (broken rotational symmetry), and chirality (broken mirror symmetry), provides a control for asymmetric transmission. In optics, optical effects arising from a single type of symmetry breaking are very common, such as birefringence (anisotropy), and circular dichroism (chirality). On the other hand, it is very rare to find optical effects that result from the interference between two or more types of symmetry breaking. In addition, we showed that asymmetric transmission can be controlled thorough adjusting the interference between two resonances with detuned resonance frequencies. Interestingly, the asymmetric transmission in amplitude vanishes for metamaterials with certain configurations, despite their low symmetry. This finding is further supported by an effective medium formulation that relates the phenomenon of asymmetric transmission to the effective electromagnetic parameters of the metamaterials.

This work is partly supported by the Engineering and Physical Sciences Council of the United Kingdom under the scheme of NSF/EPSRC Materials Network. T. Z. and S. Z. acknowledge the financial support by the European Commission under the Marie Curie Career Integration Program. J. L. acknowledges the financial support by the Research Grants Council of Hong Kong (Project No. CityU102012).

* Corresponding authors: s.zhang@bham.ac.uk, j.li@bham.ac.uk

**Appendix**

We have a planar array of the artificial atoms while the electric and magnetic dipoles generated by the artificial atoms are mainly lying on the transverse plane at $z=0$. These dipoles ($p$ and $m$) are related to the incident fields ($E_{in}$ and $B_{in}$) by

$$\begin{pmatrix} p_x \\ p_y \\ m_x \\ m_y \end{pmatrix} = [\alpha] \begin{pmatrix} E_{in,x} \\ E_{in,y} \\ B_{in,x} \\ B_{in,y} \end{pmatrix} = \begin{pmatrix} \alpha^e_{xx} & \alpha^e_{xy} & \alpha^{EH}_{xx} & \alpha^{EH}_{xy} \\ \alpha^e_{yx} & \alpha^e_{yy} & \alpha^{EH}_{yx} & \alpha^{EH}_{yy} \\ \alpha^{HE}_{xx} & \alpha^{HE}_{xy} & \alpha^m_{xx} & \alpha^m_{xy} \\ \alpha^{HE}_{yx} & \alpha^{HE}_{yy} & \alpha^m_{yx} & \alpha^{HE}_{yy} \end{pmatrix} \begin{pmatrix} E_{in,x} \\ E_{in,y} \\ B_{in,x} \\ B_{in,y} \end{pmatrix}, \quad (.1)$$

where

$$\begin{aligned} \mathbf{E}_{in} &= \mathbf{E}^+_{in} e^{ikz} + \mathbf{E}^-_{in} e^{-ikz}, \\ \mathbf{B}_{in} &= \hat{\mathbf{z}} \times \mathbf{E}^+_{in} e^{ikz} - \hat{\mathbf{z}} \times \mathbf{E}^-_{in} e^{-ikz}. \end{aligned} \quad (.2)$$

Here, we have adopted the Heaviside-Lorentz unit system. We have also assumed that we are either working at a sparse limit or working near to local atomic resonances so that we can neglect the interaction between the dipoles. In this case, the local fields to the dipolar moments are approximated by the incident fields and it makes our discussion of physics related to design of metamaterial atoms simpler. The array of the electric and magnetic dipoles radiate into free space and the total fields on the two sides become

$$\begin{aligned} \mathbf{E}(z > 0) &= \mathbf{E}^+_{out} e^{ikz} + \mathbf{E}^-_{in} e^{-ikz}, \\ \mathbf{E}(z < 0) &= \mathbf{E}^+_{in} e^{ikz} + \mathbf{E}^-_{out} e^{-ikz}, \end{aligned} \quad (.3)$$

where the output fields are given by

$$\mathbf{E}^\pm_{out} = \mathbf{E}^\pm_{in} + \frac{ik}{2A}\left(\mathbf{p} \mp \hat{\mathbf{z}} \times \mathbf{m}\right), \quad (.4)$$

with A being the area of one unit cell. If we define the scattering matrix $S_0$ (with reference plane at $z=0$) to relate the output and input fields by

$$\begin{pmatrix} \mathbf{E}^+_{out} \\ \mathbf{E}^-_{out} \end{pmatrix} = [S_0] \begin{pmatrix} \mathbf{E}^+_{in} \\ \mathbf{E}^-_{in} \end{pmatrix}, \tag{.5}$$

then by substituting Eq. ( .4), Eq. ( .1) and Eq. ( .2) into Eq. ( .5), we obtain

$$[S_0] - [I] = \frac{ik}{A}[B]^{-1}[\alpha][B] \text{ where } [B] = \begin{pmatrix} 1 & & 1 & \\ & 1 & & 1 \\ & -1 & & 1 \\ 1 & & -1 & \end{pmatrix} \tag{.6}$$

It gives us the way to extract the polarizability of the planar array of atoms. By putting Eq. ( .1) into the above equation, one can easily prove

$$\frac{t^+_{yx}}{t^-_{yx}} = \frac{(\alpha^e_{xy} - \alpha^m_{xy}) + (\alpha^{EH}_{xx} + \alpha^{EH}_{yy})}{(\alpha^e_{xy} - \alpha^m_{xy}) - (\alpha^{EH}_{xx} + \alpha^{EH}_{yy})} \tag{.7}$$

which indicates the degree of asymmetric transmission in the text.

In fact, Eq. ( .7) can be rewritten in terms of effective medium parameters if we assume the metamaterial can be homogenized into a thin slab of thickness $d$ much thinner than a wavelength, the fields are volume-averaged from Eq. ( .3) within the whole slab and become

$$\begin{aligned} \langle \mathbf{E} \rangle &= \mathbf{E}_{in} + \frac{ik\mathbf{p}}{2A}, \\ \langle \mathbf{H} \rangle &= \mathbf{H}_{in} + \frac{ik\mathbf{m}}{2A}. \end{aligned} \tag{.8}$$

By substituting Eq. ( .1) and the effective medium definition

$$\frac{1}{Ad}\begin{pmatrix} \mathbf{p} \\ \mathbf{m} \end{pmatrix} = [\chi]\begin{pmatrix} \langle \mathbf{E} \rangle \\ \langle \mathbf{H} \rangle \end{pmatrix} = \begin{pmatrix} \varepsilon_{xx}-1 & \varepsilon_{xy} & \xi_{xx} & \xi_{xy} \\ \varepsilon_{xy} & \varepsilon_{yy}-1 & \xi_{yx} & \xi_{yy} \\ -\xi_{xx} & -\xi_{yx} & \mu_{xx}-1 & \mu_{xy} \\ -\xi_{xy} & -\xi_{yy} & \mu_{xy} & \mu_{yy}-1 \end{pmatrix}\begin{pmatrix} \langle \mathbf{E} \rangle \\ \langle \mathbf{H} \rangle \end{pmatrix} \tag{.9}$$

into Eq. ( .8), we obtain the relationship between susceptibility and polarizability as

$$\frac{1}{[\chi]} = \frac{Ad}{[\alpha]} + \frac{ikd}{2} \tag{.10}$$

The susceptibility is very close to the polarizability per unit volume while the second term is regarded as the radiative correction at a small finite frequency. By substituting Eq. ( .10) into Eq. ( .6) and expanding up to the first order of $kd$, we obtain

$$t_{xy}^{+} = t_{yx}^{-} \approx \frac{1}{2}\left(\varepsilon_{xy} - \mu_{xy} - \xi_{xx} - \xi_{yy}\right)ikd$$
$$t_{yx}^{+} = t_{xy}^{-} \approx \frac{1}{2}\left(\varepsilon_{xy} - \mu_{xy} + \xi_{xx} + \xi_{yy}\right)ikd \tag{.11}$$

Therefore, we have

$$\frac{t_{yx}^{+}}{t_{yx}^{-}} \approx \frac{\left(\varepsilon_{xy} - \mu_{xy}\right) + \left(\xi_{xx} + \xi_{yy}\right)}{\left(\varepsilon_{xy} - \mu_{xy}\right) - \left(\xi_{xx} + \xi_{yy}\right)} \tag{.12}$$

This is Eq. (7) in the text.